\documentclass[a4paper,11pt]{article}
\pdfoutput=1 

\usepackage{jheppub} 

\usepackage[T1]{fontenc} 

\newcommand{\ideal}{{pluperfect }}

\begin{document}

\title{Bidirectional holographic codes and sub-AdS locality}

\author[a]{Zhao Yang}
\author[a]{Patrick Hayden}
\author[a]{Xiao-Liang Qi}
\affiliation[a]{
Stanford Institute for Theoretical Physics, \\
Physics Department, Stanford University, CA 94304-4060, USA}

\abstract{
Tensor networks implementing quantum error correcting codes have recently been used to construct toy models of holographic duality explicitly realizing some of the more puzzling features of the AdS/CFT correspondence. These models reproduce the Ryu-Takayanagi entropy formula for boundary intervals, and allow bulk operators to be mapped to the boundary in a redundant fashion. These exactly solvable, explicit models have provided valuable insight but nonetheless suffer from many deficiencies, some of which we attempt to address in this article. We propose a new class of tensor network models that subsume the earlier advances and, in addition, incorporate additional features of holographic duality, including: (1) a holographic interpretation of all boundary states, not just those in a ``code'' subspace, (2) a set of bulk states playing the role of ``classical geometries'' which reproduce the Ryu-Takayanagi formula for boundary intervals, (3) a bulk gauge symmetry analogous to diffeomorphism invariance in gravitational theories, (4) emergent bulk locality for sufficiently sparse excitations, and (5) the ability to describe geometry at sub-AdS resolutions or even flat space.
}

\maketitle

\section{Introduction} \label{sec:intro}

Holographic duality is a proposed correspondence between quantum many-body systems and quantum gravity~\cite{maldacena1998,witten1998,gubser1998}. The quantum many-body system can be considered as living on the conformal boundary of an asymptotically anti-de-Sitter (AdS) space, and the gravity theory describes the dynamics of the bulk geometry (and matter fields). A particularly interesting aspect of this duality is the essential role played by quantum entanglement. As was proposed in Ref. \cite{ryu2006}, the entanglement entropy of a boundary region corresponds to the area of the minimal geodesic bounding this region, in the same way how the black-hole entropy is related to its horizon area~\cite{hawking1971,bekenstein1973}.

The Ryu-Takayanagi (RT) formula has been investigated in many different context and proved under certain conditions. Inspired by this relation between entropy and area, the holographic duality has been proposed to be related to tensor networks~\cite{swingle2012b}, which provide a representation of many-body states by contracting a network of tensors, similar to the way a Feynman diagram is interpreted. In particular, the tensor network states forming the Multi-scale Entanglement Renormalization Ansatz (MERA)~\cite{vidal2007} were constructed to efficiently describe the ground states of critical systems, which are naturally related to holographic duality since it has the geometry of a discretized hyperbolic space. Many different efforts have been made to make the tensor network approach to holography more concrete~ \cite{hartman2013,qi2013,mollabashi2014,czech2014,evenbly2015,miyaji2015,miyaji2015b}. To define a duality one needs to have a dual description of not only particular states such as the ground state, but also quantum operators, which requires a mapping between the two theories in the whole Hilbert space. This motivated the proposal of exact holographic mapping (EHM) by one of us~\cite{qi2013}. The EHM is proposed to be a tensor network that defines a unitary mapping between two identical Hilbert spaces, those of the bulk theory and boundary theory. In other words, the bulk theory and boundary theory are two different representations of the same theory. From the point of view of the boundary theory, the bulk theory corresponds to a reorganization of the degrees of freedom, and the bulk geometry for a given boundary state is determined by the locality structure of correlation functions in the transformed bulk basis.

However, the proposal that a tensor network can define a unitary mapping rather than a state brings up new questions. An operator in the bulk will be unitarily mapped to a unique operator on the boundary. On the other hand, in the holographic duality one expects that a local operator in the low energy effective theory of the bulk can be mapped to a boundary operator that is supported on part of the boundary \cite{hubeny2012}. This boundary region is not unique, since it's not invariant under conformal transformations. Therefore there is an apparent contradiction that the bulk operator should be mapped to multiple boundary operators living in different regions on the boundary. A resolution to this paradox was proposed in Ref. \cite{almheiri2014}, which pointed out that such a non-unique correspondence can be consistently realized if the correspondence only applies to a subspace of the theory consisting of low energy states. This is related to the theory of quantum error-correction, in the sense that information corresponding to the bulk operators is encoded in the boundary theory, which has a bigger Hilbert space. Since this encoding is redundant, it is possible to recover the information from part of the boundary degrees of freedom, and the region that can be used to recover a bulk operator is not unique.

Tensor networks with such error-correction properties have been explicitly constructed in an important recent paper by F. Pastawski, B. Yoshida, D. Harlow and J. Preskill (PYHP)~\cite{pastawski2015}. By making use of tensors with particular entanglement properties, known as perfect tensors, they proposed two types of tensor networks. One is a hyperbolic tensor network that describes a state on the boundary, which they call the holographic state, for which any singly-connected region on the boundary satisfies the RT formula. The other is a tensor network with bulk and boundary indices, which defines an isometry from bulk (containing fewer uncontracted legs than the boundary) to boundary. This mapping, which PYHP call the holographic code, maps operators acting on these bulk sites to a subset of boundary operators. The mapping defines a subspace of the boundary in which the error correction properties hold, i.e. the bulk operators can be read out from different boundary regions. However, some drawbacks still exist in the PHYP proposal. In particular:
\begin{enumerate}
\item The holographic state and holographic code are two distinct tensor networks so their two key properties, namely the definition of states satisfying the RT formula and the redundant mapping of bulk to boundary operators, are not realized simultaneously.
\item The holographic code defines a mapping in a subspace of the boundary theory, which does not apply to other states on the boundary.
\item As with all other tensor network proposals so far, locality at the sub-AdS scale cannot be studied, since the network is implicitly only defined at scales larger than AdS radius.
\end{enumerate}
The objective of this paper is to provide a remedy to this collection of closely related deficiencies.
We propose a new type of tensor network, which we call a \emph{bidirectional holographic code} (BHC). Built with tensors each satisfying some internal unitarity requirements, a BHC simultaneously satisfies the following properties:
\begin{enumerate}
\item It defines an isometry from the boundary Hilbert space to the bulk. The image of this mapping is considered to be the physical Hilbert space of the bulk theory, which consists of states satisfying a gauge symmetry. When we consider the entire Hilbert space, the bulk theory is intrinsically nonlocal, with no local gauge invariant operators.
\item Some particular states of the bulk are shown to correspond to boundary states satisfying the RT formula for single intervals. These states play the role of ``classical geometries" in the bulk theory (with non-positive curvature).
\item Quantum excitations can be created by operators acting on the classical geometry states. Many operators that create a low density of excitations can be mapped to the boundary isometrically, while at high enough density the isometrical property breaks down. In other words, at a low density of excitations, the bulk sites all appear to be independent degrees of freedom, like in a local quantum field theory, while the nonlocal nature of a quantum gravity theory is revealed when highly excited states are made.
\item Distinct from previous tensor network proposals, the BHC can apply to a graph that has sub-AdS scale resolution, or even apply to a flat geometry. In a graph with sub-AdS scale resolution, there appears naively to be more bulk points than boundary points. The properties of the BHC guarantee that on one hand the actual number of bulk degrees of freedom is only proportional to those of the boundary, while on the other hand the ``low energy physics" of the bulk {\it appears} to be that of a local quantum field theory. 
\end{enumerate}

The remainder of this article is organized as follows. Section~\ref{sec:bhc} introduces the notion of a \ideal tensor, provides an explicit example and discusses the relationship between \ideal tensors and random quantum states. Section~\ref{sec:gauge} explains how to build a bidirectional holographic code by contracting a network of \ideal tensors, emphasizing the resulting bulk gauge invariance and the triviality of local bulk operators. Section~\ref{sec:emergent} then identifies a ``low-energy subspace'' of bulk states and explains how a restricted form of locality emerges in the bulk theory.

\section{Definition of the bidirectional holographic code} \label{sec:bhc}

\subsection{Building block: \ideal tensors}

\begin{figure} [h]
\begin{center}
\includegraphics[width=0.85\textwidth]{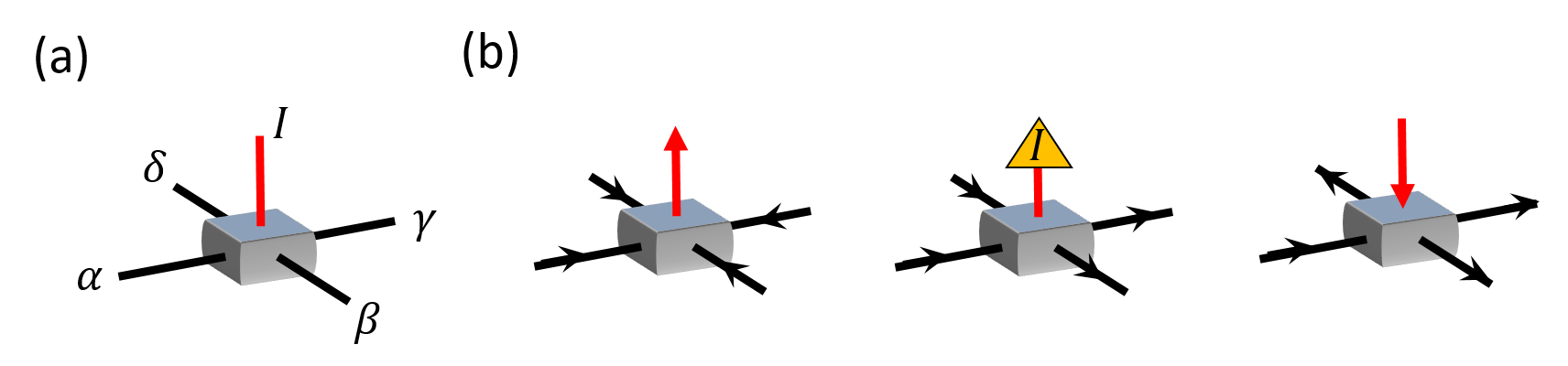}
\caption{(a) The definition of a \ideal tensor with four in-plane indices (black lines) $\alpha\beta\gamma\delta$ and one bulk index (red vertical line) $I$. (b) Illustration of the three pluperfection conditions. The arrow direction indicates that the tensor is a unitary map (up to renormalization factors) between input arrows and output arrows. The yellow triangle labeled with $I$ stands for fixing the input of the bulk index to one of the $D^2$ states $I=1,2,...,D^2$.
} \label{fig:tensor}
\end{center}
\end{figure}

A BHC will be constructed by contracting smaller building blocks, which we call \emph{\ideal tensors}, each with an odd number $(2n+1)$ of indices. These tensors reproduce the properties of the perfect tensors used as building blocks by PYHP, but must satisfy additional, more stringent conditions, hence their name. To be concrete, throughout this work we take $n=2$, although generalization to other $n$ is straightforward. For $n=2$ the tensor can be written as $T_{\alpha\beta\gamma\delta}^I$, as drawn in Fig. \ref{fig:tensor}. The four in-plane indices $\alpha,\beta,\gamma,\delta$, which run from $1,2,...,D$, will be contracted to form the tensor network. The perpendicular index $I=1,2,...,D^4$ will represent a bulk degree of freedom. A \ideal tensor is required to satisfy the following three conditions:
\begin{enumerate}
\item \label{cond1} $T$ is a unitary mapping from indices $\alpha\beta\gamma\delta$ to the bulk index $I$. As an equation, this means that
\begin{eqnarray}
T_{\alpha\beta\gamma\delta}^IT_{\alpha\beta\gamma\delta}^{J*}=\delta_{IJ}.
\end{eqnarray}
\item \label{cond2} Among the $D^4$ values of $I$, there exists a subset $I=1,2,...,D^2$, for which $T^I_{\alpha\beta\gamma\delta}$ is a four-leg perfect tensor when $I$ is fixed. In other words, $T^I_{\alpha\beta\gamma\delta}$ for fixed $I=1,2,...,D^2$ defines a unitary mapping from any two of the four indices $\alpha\beta\gamma\delta$ to the other two (up to a normalization factor). As an equation, this reads
\begin{eqnarray}
T^I_{\alpha\beta\gamma\delta}T^{I*}_{\mu\nu\gamma\delta}=\frac1{D^2}\delta_{\alpha\mu}\delta_{\beta\nu}, \text{~for~}I=1,2,...,D^2\end{eqnarray}
and other equations obtained by permutation of the four in-plane indices.
\item \label{cond3} For any one of the four in-plane indices, say $\alpha$, $T^I_{\alpha\beta\gamma\delta}$ for $I=1,2,...,D^2$ is a unitary mapping from $I\alpha$ to $\beta\gamma\delta$, again up to normalization. As an equation,
\begin{eqnarray}
T^I_{\alpha\beta\gamma\delta}T^{J*}_{\mu\beta\gamma\delta}=\frac1D\delta_{IJ}\delta_{\alpha\mu},\text{~for~}I,J=1,2,...,D^2.\label{eq:cond3}
\end{eqnarray}
\end{enumerate}
We will refer to these as the \emph{pluperfection conditions}.

Each condition will serve a specific role in the construction of the BHC. Condition \ref{cond1} facilitates the definition of a unitary holographic mapping applicable to all states on the boundary. Condition \ref{cond2} defines the particular set of ``classical geometry" states satisfying the RT formula. Condition \ref{cond3} guarantees that certain sets of operators acting on the classical geometry states are mapped to the boundary isometrically. These statements will be explained in detail as we proceed. Before doing that, however, we would like to give an explicit example to demonstrate that \ideal tensors do exist, and also discuss why it is natural to consider them.

\subsection{Explicit construction of a \ideal tensor}

To demonstrate that \ideal tensors do exist, we present an explicit example. We start from a particular $[403]_3$ stabilizer code , which is a four-leg tensor $R_{\alpha\beta\gamma\delta}$ with each leg having dimension $D=3$~\cite{gottesman97}. We can view this as a quantum state $|\Psi\rangle=R_{\alpha\beta\gamma\delta}|\alpha\rangle|\beta\rangle|\gamma\rangle|\delta\rangle$. This state is defined by the equation $S_i|\psi\rangle=|\psi\rangle$, for $i=1,2,3,4$, with the stabilizer operators $S_i$ defined to be
\begin{eqnarray}
\left\{S_1,S_2,S_3,S_4\right\}&=&\left\{ZZZI,~ZZ^{-1}IZ,~XXXI,~XX^{-1}IX\right\}\label{stabilizers}\\
\text{with~}X&=&\left(\begin{array}{ccc}0&1&0\\0&0&1\\1&0&0\end{array}\right),~Z=\left(\begin{array}{ccc}1&0&0\\0&e^{2\pi i/3}&0\\0&0&e^{4\pi i/3}\end{array}\right)\nonumber
\end{eqnarray}
Here $ZZZI$ denotes a direct product of $Z$ operators acting on the first three legs and identity operator $I$ acting on the last leg. Other terms are defined similarly.  By construction, this four-leg tensor is a perfect tensor. This motivates us to take $T^{I=0}_{\alpha\beta\gamma\delta}=R_{\alpha\beta\gamma\delta}$ and try to construct the other components of $T$. Since local unitary transformations do not alter the entanglement properties, it is easy to see that for any four local unitary operators $g_1,g_2,g_3,g_4\in SU(D)$ acting on the four legs, $g_1g_2g_3g_4|\psi\rangle$ will still be a perfect tensor. Therefore if we find $D^2$ orthogonal states, all in the form of $g_1g_2g_3g_4|\psi\rangle$, pluperfection Condition \ref{cond2} is automatically satisfied. Then we can do further selection to find a choice to satisfy Condition \ref{cond3}. Once Condition \ref{cond3} is  satisfied, by contraction of $\alpha$ and $\mu$ in Eq. (\ref{eq:cond3}) one can see that the unitarity condition \ref{cond1} is already satisfied for the subset of states $I,J=1,2,...,D^2$. Therefore it's trivial to satisfy Condition \ref{cond1} by adding orthogonal vectors to these $D^2$ states to make a complete $D^4$ dimensional basis.

Following this line of thought, and making use of the fact that commutation relations between the generalized Pauli operators (products of $X$ and $Z$ defined in Eq. (\ref{stabilizers})) are easy to compute, it is easy to make a numerical search and find \ideal tensors satisfying our requirements. (More details about the code and search are given in Appendix \ref{app:stabilizer}.) In the following we write down one particular option. We introduce two operators
\begin{eqnarray}
A=XXZZ^{-1},~B=ZZ^{-1}XX\label{choiceofAB}
\end{eqnarray}
and define the $D^2=9$ states corresponding to $T^I_{\alpha\beta\gamma\delta}|\alpha\rangle|\beta\rangle|\gamma\rangle|\delta\rangle$ as
\begin{eqnarray}
|nm\rangle=A^nB^m|\psi\rangle,~n,m=0,1,2
\end{eqnarray}
Since for any $n,m$, $A^nB^m$ is a direct product of four local unitary operators acting on the four qutrits, Condition \ref{cond2} is automatically satisfied. To verify Condition \ref{cond3}, we express it in the following equivalent form:
\begin{itemize}
\item For any single site operator $O$ acting on any one of the four sites, $\langle nm|O|kl\rangle=\frac1D\delta_{nk}\delta_{ml}{\rm tr}(O)$.
\end{itemize}
For the qutrit system, one just needs to verify this condition for $O=X^aZ^b$, for $a,b=0,1,2$. Since  generalized Pauli operators commute up to phase, the condition $\langle nm|O|kl\rangle=\frac1D\delta_{nk}\delta_{ml}{\rm tr}(O)$ is equivalent to the statement that $\left[A^{n-k}B^{m-l}\left(X^aZ^b\right),S_i\right]=0$ holds for all $i$ only if $n-k=0,~m-l=0$ and $a=b=0$. This statement can be explicitly verified for the choice of $A,B$ in Eq. (\ref{choiceofAB}), establishing that the $9$ states $|nm\rangle$ define a tensor satisfying Conditions \ref{cond2} and \ref{cond3}. To define a tensor $T^{I}_{\alpha\beta\gamma\delta}$ with $D^4$ dimensional index $I$ we simply need to introduce a $D^4=81$ dimensional basis in the 4-qutrit Hilbert space which contains $|nm\rangle$ as $9$ of the basis vectors.


\subsection{Pluperfect tensors as idealized random states}

While the somewhat intricate construction of the previous subsection would suggest that \ideal tensors are very special, they can actually be thought of as idealized versions of random tensors in high dimension $D$.
As first observed by Lubkin~\cite{L78}, states chosen at random according to the unitarily invariant measure will be highly entangled. The average entropy of two sites in a four-qudit Hilbert space will be
\begin{equation}
S
= \sum_{j=D^2+1}^{D^4} \frac{1}{j} - \frac{D^2-1}{2D^2}
\longrightarrow 2 \log D - \frac{1}{2}
\end{equation}
in the limit of large $D$~\cite{P93,FK94}, a result known as Page's theorem. In entropic terms, therefore, a random four-qudit tensor will be maximally entangled up to a small dimension-independent deficit.

If that deficit were exactly zero, the tensor would be maximally entangled across all evenly weighted cuts, which is the definition of a perfect tensor. Thus, as discussed by PYHP in their Appendix A.3, a perfect tensor is an idealized version of a random tensor~\cite{pastawski2015}. Moreover, interpreted as linear maps, random tensors will have maximal rank except on a set of measure zero, allowing operators to be pushed through them with the caveat that unitarity will not be preserved.

In the same spirit, pluperfect tensors are also idealizations of random tensors. If we randomly select $D^2$ orthogonal states \{$|I\rangle : I = 1,\ldots, D^2$\}, again according to the unitarily invariant measure, then each of them will automatically satisfy Page's theorem, so that Condition \ref{cond2} will be approximately satisfied in the large $D$ limit. To check Condition \ref{cond3} we consider a single site operator $O_1$, and calculate its matrix element between two different states $|I\rangle$ and $|J\rangle$. By doing a Haar average we obtain
\begin{eqnarray} \label{eqn:off-diag}
\left|\langle I|O_1|J\rangle\right|^2=\frac1{D^8-1}\left[D^3{\rm tr}\left(O_1O_1^\dagger\right)-D^2{\rm tr}\left(O_1\right)^2\right]=O\left(D^{-5}\right).
\end{eqnarray}
Therefore in large $D$ limit, Condition \ref{cond3} is also asymptotically approximately satisfied, and thus we conclude that \ideal tensors are indeed idealized versions of $D^2$ Haar-random orthogonal states in $D^4$-dimensional Hilbert space in the large $D$ limit.

There is an important caveat to this discussion, however, which is that the weak sense in which random tensors approximate \ideal tensors is not obviously strong enough to allow their direct substitution for \ideal tensors in most of the arguments of this article. It would be interesting to study the propagation of errors in a random tensor network to see whether the structure identified in this article survives.


Another source of intuition for Condition \ref{cond3} is to compare it to the Eigenstate Thermalization Hypothesis (ETH), which proposes conditions conjectured to hold widely in quantum chaotic systems, under which the expectation values of observables will converge to their microcanonical averages~\cite{deutsch1991quantum,srednicki1994chaos}. An observable $O$ is said to satisfy the ETH  if: (1) in a basis $|E_\alpha\rangle$ of energy eigenstates, the diagonal entries $O_{\alpha \alpha} = \langle E_\alpha | O | E_\alpha \rangle$ change only insignificantly in a microcanonical energy window, while (2) the off-diagonal elements $O_{\alpha \beta} = \langle E_\alpha | O | E_\beta \rangle$ for $\alpha \neq \beta$ are exponentially small in the number of degrees of freedom. Since Condition~\ref{cond3} implies that $\langle I | O_1 | J \rangle = 0$ whenever $I \neq J$, every single-site observable satisfies a strengthened \emph{exact} form of the second part of the ETH. The $D^2$ orthogonal states $\{ |I\rangle \}$ of a pluperfect tensor therefore behave like idealized eigenstates of a quantum chaotic system with respect to single-site operators.

\section{Gauge invariance and physical operators} \label{sec:gauge}

\begin{figure} [h]
\begin{center}
\includegraphics[width=0.45\textwidth]{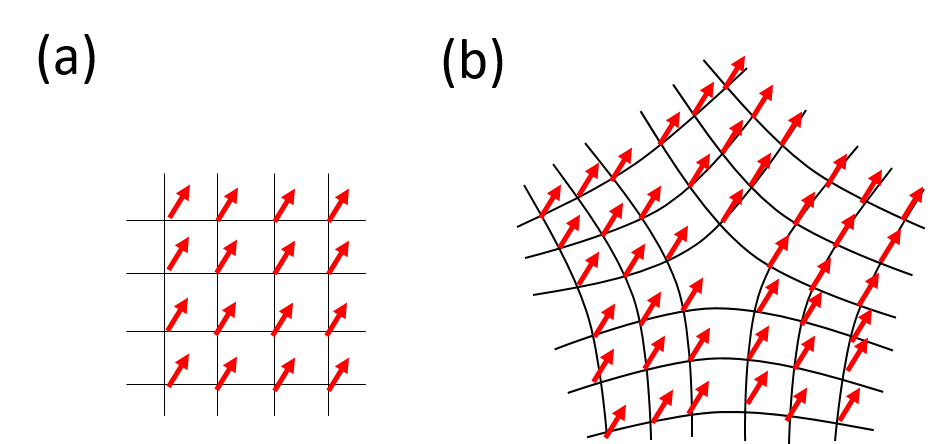}
\caption{Two examples of tensor networks made by contracting \ideal tensors.
} \label{fig:networks}
\end{center}
\end{figure}

With the \ideal tensors defined, we can contract the in-plane indices of this tensor and build tensor networks with different geometry. Examples of tensor networks are shown in Fig. \ref{fig:networks}. In the following two sections, we will study the properties of such tensor networks. The discussion in this section will focus on the properties of the network that only rely on Condition \ref{cond1}, with the consequences of Conditions \ref{cond2} and \ref{cond3} postponed to the next section.

\begin{figure} [h]
\begin{center}
\includegraphics[width=0.9\textwidth]{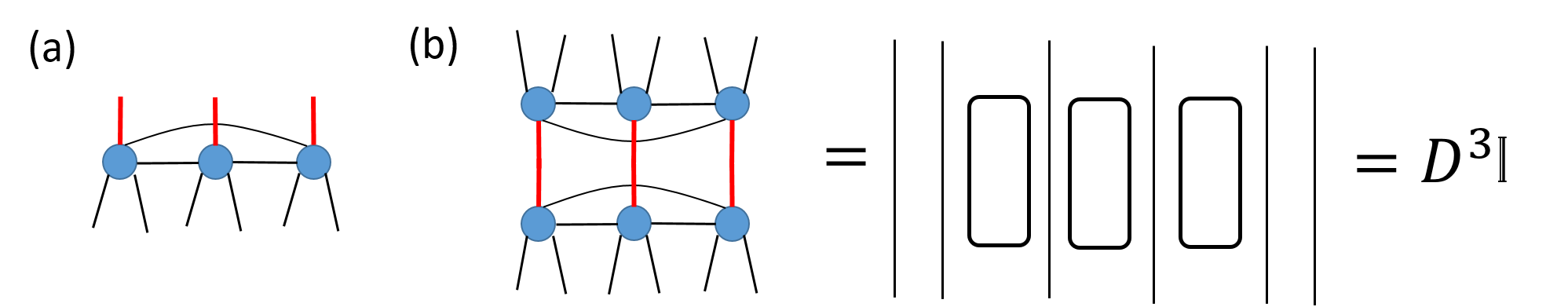}
\caption{(a) A simple example of a tensor network with three \ideal tensors. (b) Contraction of the bulk indices which proves that the mapping $M$ is an isometry, after proper normalization.
} \label{fig:isometry}
\end{center}
\end{figure}

When we consider a network with $V$ vertices. In the $4V$ in-plane lines of these tensors, $N$ pairs of them are contracted, leaving $P=4V-2N$ boundary legs. This tensor network can be viewed as a linear map
\begin{eqnarray}
M:~\mathbb{H}_\partial\longrightarrow \mathbb{H}_A
\end{eqnarray}
with $\mathbb{H}_\partial$ the $D^P$ dimensional Hilbert space of boundary indices, and $\mathbb{H}_A$ the (bigger) $D^{4V}$ dimensional Hilbert space of bulk indices. For later convenience, we define the linear map $M$ by contracting all the internal lines and multiplying a normalization factor $D^{-N/2}$. It is easy to prove that this map (with proper normalization) is an isometry. As is illustrated in Fig. \ref{fig:isometry} for a simple example, Condition \ref{cond1} means that $T^I_{\alpha\beta\gamma\delta}T^{I*}_{\mu\nu\sigma\tau}=\delta_{\alpha\mu}\delta_{\beta\nu}\delta_{\gamma\sigma}\delta_{\delta\tau}$. Using this equation one can prove
\begin{eqnarray}
M^\dagger M=\mathbb{I}
\end{eqnarray}
with $\mathbb{I}$ the identity operator in the boundary Hilbert space. We would like to treat the boundary theory (a quantum many-body system with no gravity) as our starting point since that is better understood. The physical bulk state Hilbert space $\mathbb{H}_b$ is defined as the image of the mapping $\mathbb{H}_b=M(\mathbb{H}_\partial)$, which is a subspace of the naive bulk Hilbert space $\mathbb{H}_A$. In other words, the physical bulk states are defined by the equation
\begin{eqnarray}
MM^\dagger|\Psi\rangle=|\Psi\rangle.
\end{eqnarray}

To obtain a more explicit understanding of this physical state condition, we now show that the physical states are invariant with respect to a  ``generalized gauge symmetry". Consider four arbitrary $SU(D)$ transformations $g_1,g_2,g_3,g_4$, which act on the four in-plane qudits of a \ideal tensor. Since the \ideal tensor $T$ is unitary by Condition \ref{cond1}, the transformation $g_1\otimes g_2\otimes g_3\otimes g_4$ can be mapped to the bulk index by conjugation, which defines a transformation
\begin{eqnarray}
W\left(g_1,g_2,g_3,g_4\right)=T\left(g_1\otimes g_2\otimes g_3\otimes g_4\right)T^\dagger.\label{gaugepush}
\end{eqnarray}
Here we have denote the unitary mapping from $\alpha\beta\gamma\delta$ to $I$ defined by $T^I_{\alpha\beta\gamma\delta}$ also as $T$. By definition, the action of $W$ on the bulk index is equivalent to the action of $g_1\otimes g_2\otimes g_3\otimes g_4$ on the in-plane indices.

In a tensor network obtained by connecting \ideal tensors, there is an obvious gauge redundancy corresponding to each internal line that is contracted. Denote by $x$ and $y$ two sites in a tensor network that are connected by one internal line. The contraction of this line leads to $T^I_{\alpha\beta\gamma\delta}T^J_{\alpha\mu\nu\tau}$ (where we have assumed the contracted index to be the first one in each tensor). For an $SU(D)$ matrix $g$ with indices $g_{\alpha\beta}$, we have $g_{\alpha\pi}T^I_{\pi\beta\gamma\delta}g^*_{\alpha\epsilon}T^J_{\epsilon\mu\nu\tau}=T^I_{\pi\beta\gamma\delta}T^J_{\pi\mu\nu\tau}$, so that acting with $g$ and $g^*$ on the first indices before the contraction leads to the same state. If we write $g_{xy}$ for the $SU(D)$ gauge transformation associated with each link, with the convention that $g_{yx}=g_{xy}^*$, then the $N$ internal lines define an $SU(D)^N$ gauge group. Such gauge transformations can then be pushed to the bulk by Eq. (\ref{gaugepush}). Each choice of $g_{xy}\in SU(D)^N$ defines a set of operators $W_x\left(g_{xy_1},g_{xy_2},g_{xy_3},g_{xy_4}\right)$ acting on each bulk site (with $y_{1,2,3,4}$ the four neighbors of $x$). The physical state condition can now be written in a local form:
\begin{eqnarray}
W_x\left(g_{xy_1},g_{xy_2},g_{xy_3},g_{xy_4}\right)|\Psi\rangle=|\Psi\rangle,~\text{for arbitrary}~\left\{g_{xy}\right\}.
\end{eqnarray}

Such a gauge symmetry is different from the usual form found in gauge theory, where the gauge transformations are defined on sites and the gauge vector potentials are defined on lines, which are transformed by the gauge transformations at the adjacent sites. In contrast, in the current case the gauge transformations are defined on links, and the $W_x$ acting on the bulk degrees of freedom depend on the four links attached to site $x$. An important consequence of this difference is that ordinary gauge theories have gauge invariant operators that are Wilson loops, while {\it there are no gauge invariant local operators in the bulk physical Hilbert space} in our theory.

To be more precise, we denote the set of bulk sites that are not directly adjacent to the boundary as the interior. We have the following statement:

\begin{itemize}
\item Any operator $O$ in the Hilbert space $\mathbb{H}_A$ which acts only in the interior is mapped to a trivial operator ({\it i.e.} proportional to the identity operator) on the boundary.
\end{itemize}

\begin{figure} [h]
\begin{center}
\includegraphics[width=0.9\textwidth]{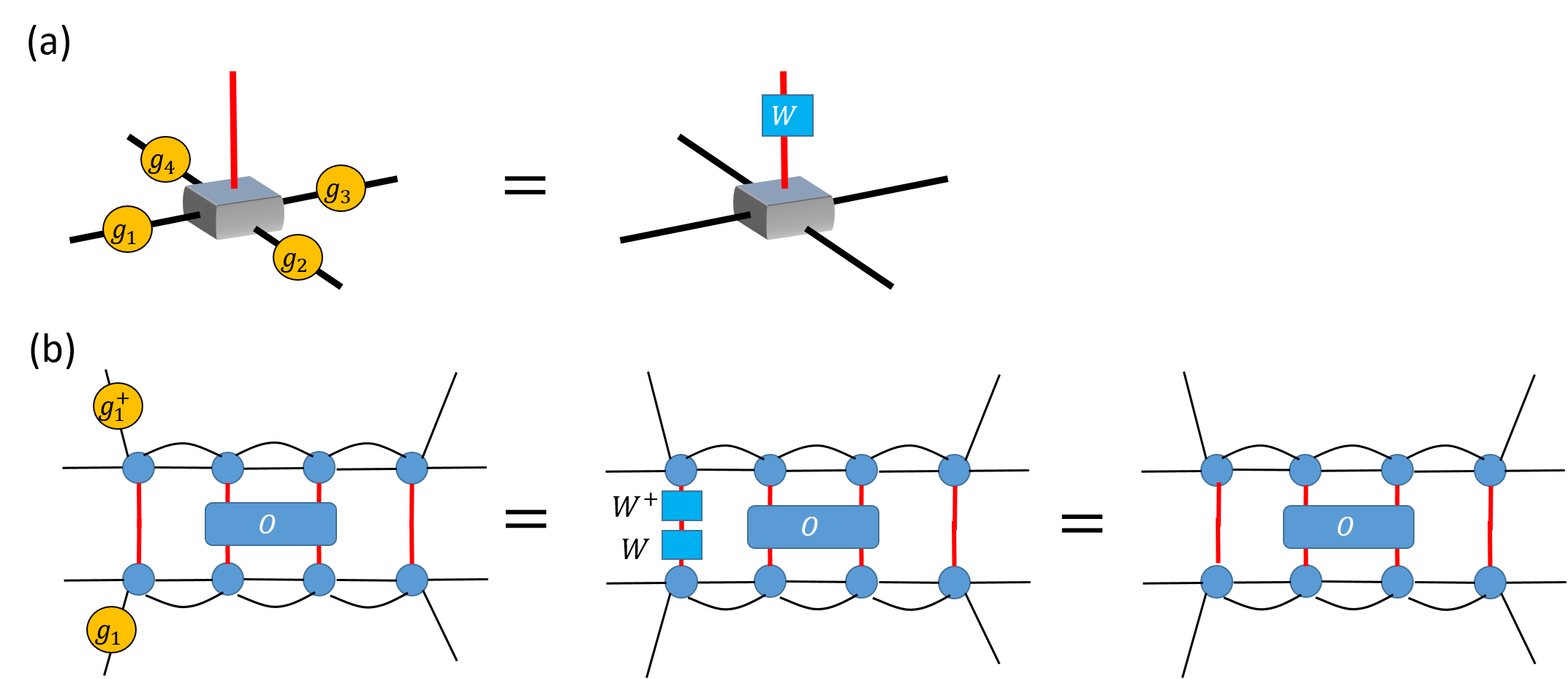}
\caption{(a) The gauge transformation on in-plane indices that is equivalent to a bulk operator $W$. (b) For a simple network, an illustration of why the boundary projection of an arbitrary operator in the interior $M^\dagger OM$ commutes with a boundary ${\rm SU(D)}$ transformation $g_1$.
} \label{fig:gauge}
\end{center}
\end{figure}

The operator $O$ is mapped to a boundary operator $M^\dagger OM$. To prove that this operator is trivial, we can consider a unitary transformation $g_{xa}$ acting on a boundary leg. This unitary transformation can then be pushed to the bulk becoming some $W_x$ (with all other gauge transformations trivial). Since $O$ does not act on the boundary site $x$, $W_x$ cancels with $W_x^{-1}$, and we obtain $\left[g_{xa},M^\dagger OM\right]=0$. Since this equation applies to all boundary legs for all arbitrary $g_{xa}$, we conclude that $M^\dagger OM\propto \mathbb{I}$. A graphic derivation is shown in Fig. \ref{fig:gauge} (b).

BHC gauge invariance can be viewed as an analog of general covariance in the tensor network language. Due to our Condition \ref{cond1}, the \ideal tensor $T^I_{\alpha\beta\gamma\delta}$ is a unitary map from bulk index $I$ to the in-plane indices. Therefore, if a bulk site is in a single-site pure state with the wavefunction $\psi_I$, by varying $\psi_I$ one can realize an arbitrary four-leg tensor by $t_{\alpha\beta\gamma\delta}=T^I_{\alpha\beta\gamma\delta}\psi_I$. In other words, the naive $D^{4V}$ dimensional Hilbert space is precisely the space of all tensor network states (defined on the given graph). Each direct product state in the bulk $\prod_x|\psi_x\rangle$ is mapped to a tensor network state with each tensor determined, while the entangled states in the bulk can be interpreted as linear superpositions of different tensor networks. A physical state on the boundary can have many different tensor network representations, and the equivalence between them is reflected in the gauge symmetry. If we compare this with a continuum gravity theory, it is natural to draw the analogy that different metrics in the bulk that are related by diffeomorphism (which acts trivially on the boundary) are equivalent descriptions of the same boundary theory. BHC gauge invariance can thus be viewed as an analog of diffeomorphism invariance, but the former is potentially an even bigger gauge symmetry.

\section{Emergent locality in the ``low-energy subspace"} \label{sec:emergent}

The gauge invariance property discussed in the previous section only relies on pluperfection Condition \ref{cond1}, which means that it applies to much more generic tensor networks than BHC's. In fact, tensors satisfying Condition \ref{cond1} can be viewed as special cases of injective projected entangled pair states (PEPS)\cite{verstraete2004,perez2008}. For an injective PEPS, an isometry from boundary indices to bulk indices can be defined for regions exceeding certain critical size. In this sense, any injective PEPS defined on a big enough graph (such that it's injective) defines an isometry from the boundary to the bulk, which can thus be viewed as a ``holographic mapping", with gauge invariance properties similar to those discussed in the previous section.

What then is special for the holographic mapping we are seeking, which makes the bulk theory qualify as a quantum gravity theory? Although a complete answer to that question may be difficult, a necessary condition is that the bulk theory should have emergent locality in a certain subspace of states, usually called the ``low energy subspace." A generic state in a quantum gravity theory contains black holes  and strong quantum fluctuations of geometry, with the consequence that locality is not defined. This is consistent with the fact that there are no local gauge invariant operators. However, the interesting quantum gravity theories are those with a classical limit, which describes weak fluctuations (gravitons and other matter fields) around a classical geometry. Restricted to the low energy excitations around this semiclassical vacuum state, the theory looks like a $(d+1)$-dimensional quantum field theory, although (according to the holographic principle) its actual number of degrees of freedom is proportional to the $d$-dimensional boundary area. In this section, we will show how such requirements can be satisfied by the particular tensor choice in BHC. We first define a set of states corresponding to classical geometries, with entanglement entropy satisfying the RT formula (by applying the PYHP results), and then discuss how operators acting on these states are consistent with emergent locality.

\subsection{Classical geometry states}

\begin{figure} [h]
\begin{center}
\includegraphics[width=0.8\textwidth]{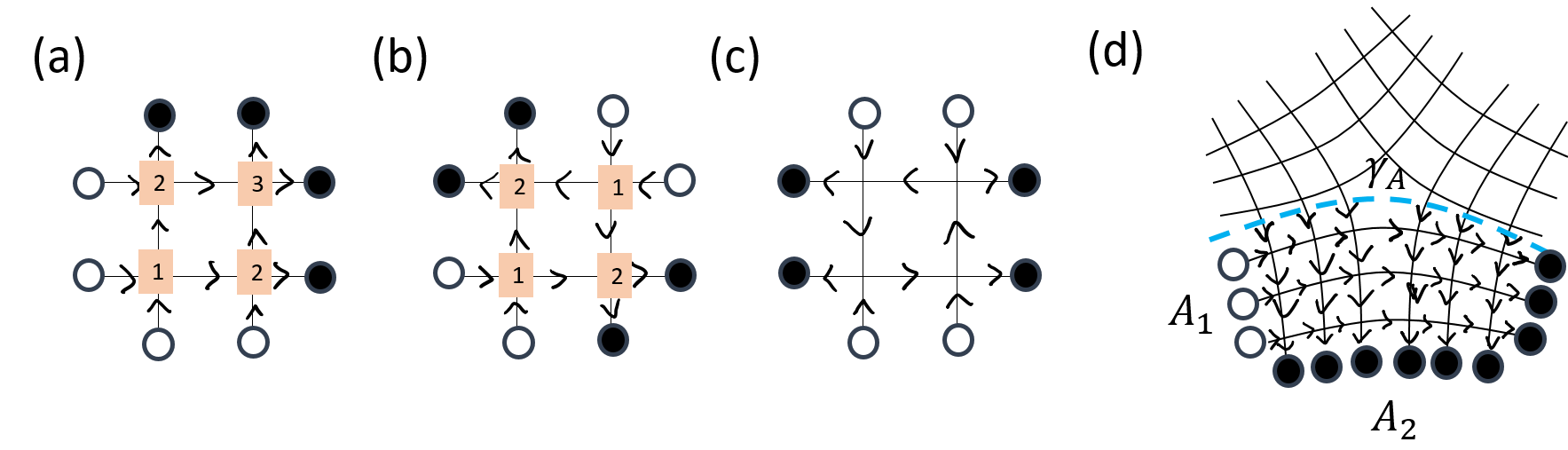}
\caption{(a) and (b) Allowed arrow assignments that define a unitary mapping from the input sites (hollow circles) to the output sites (solid circles). The number on each vertex defines the ordering of the sites. (c) illustrates a disallowed arrow configuration with a closed loop of arrows. With any order assignment, there will be arrows going against the ordering (from larger number to smaller number), so that this network does not define a unitary map from the input sites to the output sites. (d) Arrow drawing on part of a graph with non-positive curvature. $A_1$ and $A_2$ are boundary regions labeled by the hollow circles and the solid circles, respectively. The region $A_1\cup A_2=A$ bounds a geodesic line $\gamma_A$. The arrow drawing in the region between $\gamma_A$ and $A$ (the ``entanglement wedge") defines a unitary mapping from $\gamma_A\cup A_1$ to $A_2$, which thus defines an isometry from $\gamma_A$ to $A$.
} \label{fig:arrowdrawing}
\end{center}
\end{figure}

We consider a network of \ideal tensors, and a particular product state $\prod_{x}|n_x\rangle$ in the bulk. Here $n_x=1,2,...,D^2$ corresponds to setting the bulk index to a fixed value $I=n_x$. According to pluperfection Condition \ref{cond2}, the boundary state corresponding to this state (denoted by $M^\dagger \prod_{x}|n_x\rangle$) is a tensor network state consisting of contracted $4$-leg perfect tensors. If the tensor network is flat or negatively curved, we can immediately conclude from the results of PYHP~\cite{pastawski2015} that the RT formula applies to any singly-connected boundary region. To make our discussion self-contained, here we sketch the proof of this statement. Due to pluperfection Condition \ref{cond2}, we can draw arrows on a four-leg tensor, as shown in Fig. \ref{fig:tensor} (b), by choosing any two legs as inputs and the other two as outputs. A unitary map is defined between the input and output legs. Now in a network of such tensors, one can draw arrows on each link, including the boundary links and the internal links, as shown in Fig. \ref{fig:arrowdrawing}. If one can order all vertices in the network (shown in the numbers in Fig. \ref{fig:arrowdrawing} (a) and (b)), such that arrows only go from earlier layers to later ones, then a unitary mapping is defined by multiplication of the unitary matrices corresponding to each tensor following the assigned order. (More explicitly, the multiplication is defined by taking a direct product of all matrices in the each layer, and then multiplying them from right to left following the ordering.) Not all arrow assignments satisfy this requirement. An example for which such an assignment is impossible is shown in Fig. \ref{fig:arrowdrawing} (c).

Now we consider an arrow assignment for part of a graph, as shown in Fig. \ref{fig:arrowdrawing} (d). For a singly connected boundary region $A$ ($A=A_1\cup A_2$ in this figure), there is a geodesic line $\gamma_A$ bounding this region. The bulk region enclosed by $A$ and $\gamma_A$ is the ``entanglement wedge". If an arrow assignment can be drawn in the entanglement wedge, such that all sites at $\gamma_A$ are input sites (see Fig. \ref{fig:arrowdrawing} (d) as an example), a unitary mapping is defined from $\gamma_A$ and part of the boundary ($A_1$ in the figure) to the remainder of the boundary ($A_2$). Therefore an isometry is defined from $\gamma_A$ to $A$. If the same is true for the complement of $A$, the whole tensor network can be viewed as two isometries $V(A)\otimes V(\bar{A})$ acting on the EPR pairs represented by lines crossing $\gamma_A$. Denote the number of links crossed by $\gamma_A$ as $\left|\gamma_A\right|$, we have $S_A=S_{\bar{A}}=\left|\gamma_A\right|\log D$. This is the RT formula for this tensor network. A key result proved by PYHP is that such an arrow drawing can always be found for a singly connected region on the boundary, provided the bulk has no positive curvature. To be more precise, the non-positive curvature condition requires that the geodesic distance between two points $x,y$ on the dual graph, as a function of position of $y$ for a fixed $x$, has no maximum except at the boundary.

An alternative way to understand the isometries from the geodesic to the boundary is to consider a step-by-step deformation of the geodesic $\gamma_A$ to the boundary $A$, where each step is defined by moving the curve across one vertex. An isometry can be defined if the length of the curve does not decrease during each step of the deformation. For geometries with positive curvature, the geodesic may have to increase and then decrease its length before reaching the boundary. 

Although PYHP's work focused on hyperbolic geometries, the discussion applies to flat space, such as the square lattice as illustrated in Fig. \ref{fig:networks} (a). The RT formula indicates that {\it any two adjacent edges of the square are in a maximally mixed state} since the boundary itself is a geodesic. Therefore compared with the hyperbolic network, a flat-space tensor network represents a highly entangled state.

Another comment we would like to make is that the classical geometry states do not form a linear space. In other words, a linear superposition of the product states $\prod_x\left|n_x\right\rangle$ generically do not satisfy the RT formula.

When we consider a boundary region that contains multiple intervals, there are generally several different geodesic surfaces bounding them. In this case, arrow drawing fails to define isometries from the geodesic to the boundary and its complement. Two different sets of geodesics are isometrically mapped to the boundary region $A$ and its complement $\bar{A}$. Therefore the RT formula cannot be generically proved. However, the arrow drawing is a sufficient but not necessary condition for defining isometries, so that it is still possible that the RT formula holds for certain tensor networks. \footnote{An example of tensor networks satisfying the RT formula for multiple regions is a rectangle with a square lattice of the $[403]_3$ qutrit perfect tensor. The two parallel shorter edges of the rectangle is in a maximally mixed state, as is required by the RT formula. \cite{hosur2015}.} 


\subsection{Bulk ``local" operators and the causal wedge}

Now we consider operators acting on the bulk degrees of freedom. As discussed in Section~\ref{sec:gauge}, there are no local gauge invariant operators in the bulk theory. However some emergent ``local operators" can be defined if we have a classical geometry state as a reference. Consider a classical state, say $|G\rangle\equiv\prod_x|n_x=1\rangle$, as the classical vacuum. A bulk excitation can be created by applying a local operator $\phi_x$ to one of the bulk sites. For a bulk site that is not adjacent to the boundary, we know that the mapping of $\phi_x$ to the boundary gives $M^\dagger \phi_xM\propto\mathbb{I}$ which is a trivial operator. However, since we are considering the action of this operator to the vacuum state $|G\rangle$, we can map this operator to the boundary {\it while knowing all other sites are at $n_y=1$}. In other words, we can consider the boundary operator
\begin{eqnarray}
O_x=M^\dagger \phi_x\otimes \prod_{y\neq x}P_yM
\end{eqnarray}
with $P_y=|n_y=1\rangle\langle n_y=1|$ the projector to the $n_y=1$ vacuum state.

\begin{figure} [h]
\begin{center}
\includegraphics[width=0.8\textwidth]{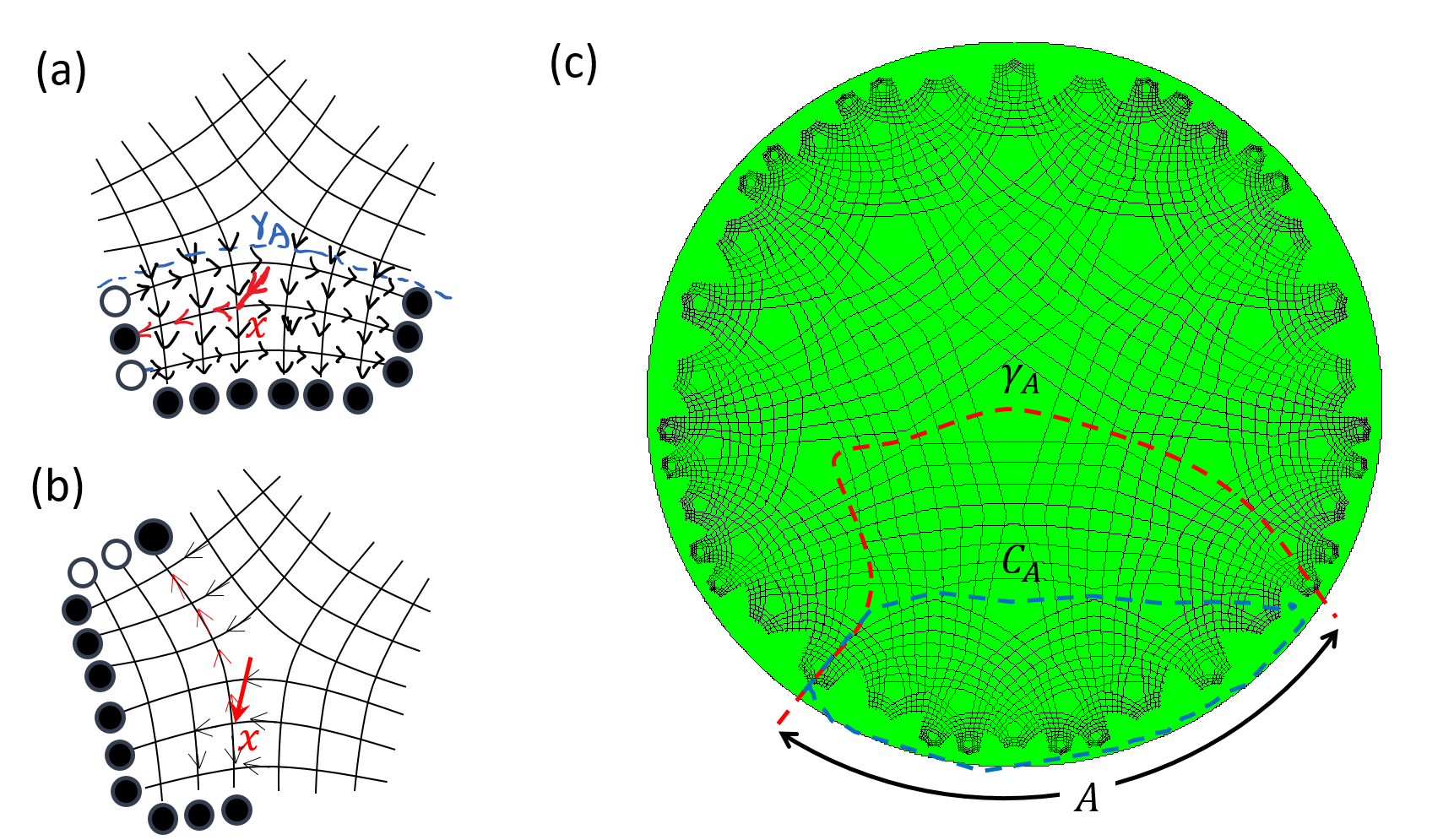}
\caption{(a) An arrow assignment to part of the graph which defines an isometry from a single bulk site $x$ to part of the boundary. The arrows define a unitary map from $\{x\}\cup \gamma_A\cup A_1$ to $A_2$, where $\gamma_A$ is the geodesic, $A_1$ is the set of the sites labeled by hollow circles and $A_2$ is the set of those with solid circles. (b) A different arrow drawing which defines an isometry from site $x$ to another boundary region. (c) For a boundary region $A$, $\gamma_A$ (red dashed curve) is the outermost geodesic bounding $A$. Not all bulk sites in the entanglement wedge enclosed by $A$ and $\gamma_A$ can be mapped to the boundary region $A$ isometrically. The set of sites with such isometries form the smaller ``causal wedge" $C_A$ enclosed by the blue dashed line.
} \label{fig:code}
\end{center}
\end{figure}

Now we would like to make use of pluperfection Condition \ref{cond3}
to show that the mapping from $\phi_x$ to $O_x$ is isometric. When
all other sites $y\neq x$ are fixed into state $|1\rangle$, the
tensor network $\prod_{y\neq x}\langle n_y=1|M$ defines a linear
map from the site $x$ to all boundary sites. If it's an isometry,
this map defines the ``encoding map" of a holographic code, which
encode the ``logical qudit" at bulk site $x$ into boundary
qudits. Similar to the discussion in the previous subsection, we
can use arrow assignments to define the isometry. On site $x$ we
draw an incoming arrow from the bulk leg and one of the in-plane
legs, as shown in Fig. \ref{fig:tensor} (b). A unitary mapping
is defined from the input arrows to the output arrows (remember
that the bulk site has dimension $D^2$). Then for other sites in
this system, the arrow assignment is two-in-two-out, the same as
in the holographic state. One also needs to make sure that the
arrows do not form any loops. Such an arrow assignment is shown in
Fig. \ref{fig:code} (a). This figure also illustrate why it is
possible to define an isometry from a bulk site to part of the
boundary. Similar to the discussion of the RT formula earlier, if an arrow
assignment can define a unitary mapping in which all links
acrossing the geodesic $\gamma_A$ are used as input, then an
isometry is defined from the site $x$ to the output sites (solid
circles in Fig. \ref{fig:code} (a)). Therefore each operator at
site $x$ can be mapped to a boundary operator supported on these
boundary sites. A different arrow drawing can define an isometry
from the site $x$ to a different boundary region, as shown in
Fig. \ref{fig:code} (b).

By analogy with the terminology used in AdS/CFT~\cite{hubeny2012}, we
can define the set of all bulk sites that can be mapped
isometrically to a boundary region $A$ as the causal wedge of
$A$. In the example in Fig. \ref{fig:code} (a), the causal wedge
is the same as the entanglement wedge. In general these two
regions are distinct. As shown in Fig. \ref{fig:code} (c), the
causal wedge is generically a subset of the entanglement
wedge. Because it is impossible to define an isometric mapping,
from the bulk sites enclosed by $\gamma_A$ but not by $C_A$, along
with $\gamma_A$, to the boundary $A$.

In fact, if there are multiple geodesics homologous to the region
$A$, only the one that encloses the minimal number of bulk points
is the candidate for the boundary of the causal wedge $C_A$. The
proof is as follows. Assume there are two geodesics $\gamma_{A1}$,
$\gamma_{A2}$, and the bulk sites enclosed by $\gamma_{A2}\cap A$
form a subset of those enclosed by $\gamma_{A1}\cap A$. We claim
that operators acting on bulk points $\{x|x \in \{\gamma_{A1}\cap
A\}, x\notin \{\gamma_{A2}\cap A\}\}$ cannot be reconstructed
isometrically out of region $A$. The reason is the following. By
construction, an arrow drawing that defines an isometry from
$\gamma_{A1}$ to the boundary has all arrows pointing to $A$ at
$\gamma_{A1}$. Since $\gamma_{A2}$ is of the same length as
$\gamma_{A1}$, the directions of the bonds crossing $\gamma_{A2}$
must also point towards $A$, and the tensor network enclosed by
$\gamma_{A1}\cap \gamma_{A2}$ must be a unitary mapping from
$\gamma_{A1}$ to $\gamma_{A2}$. Thus if we now include the bulk
point $x$ as an additional input, it will be impossible to define
an isometry of $\gamma_{A1}\cup \{x\}$ to $\gamma_{A2}$. Therefore
the bulk operators acting on $x$ cannot be constructed
isometrically on region $A$, and $\gamma_{A1}$ cannot be the
boundary of the causal wedge. It seems reasonable to conjecture
that the geodesic that encloses smallest number of bulk sites is
always the boundary of the causal wedge, but we haven't been able
to rigorously prove that. 

\subsection{Kinematics of the bulk holographic theory}

The encoding map discussed in the previous subsection can be generalized to multiple bulk sites. If an isometry can be defined from a set of bulk sites to the boundary, this isometry defines a ``dictionary" which maps arbitrary bulk operators on these sites to the boundary, preserving all commutation relations. Since all other bulk sites that are outside this set are also used in this encoding map, the dictionary defined this way depends on the background geometry. This is expected, since it is not possible to define bulk local operators without referring to the background.

\begin{figure} [h]
\begin{center}
\includegraphics[width=0.85\textwidth]{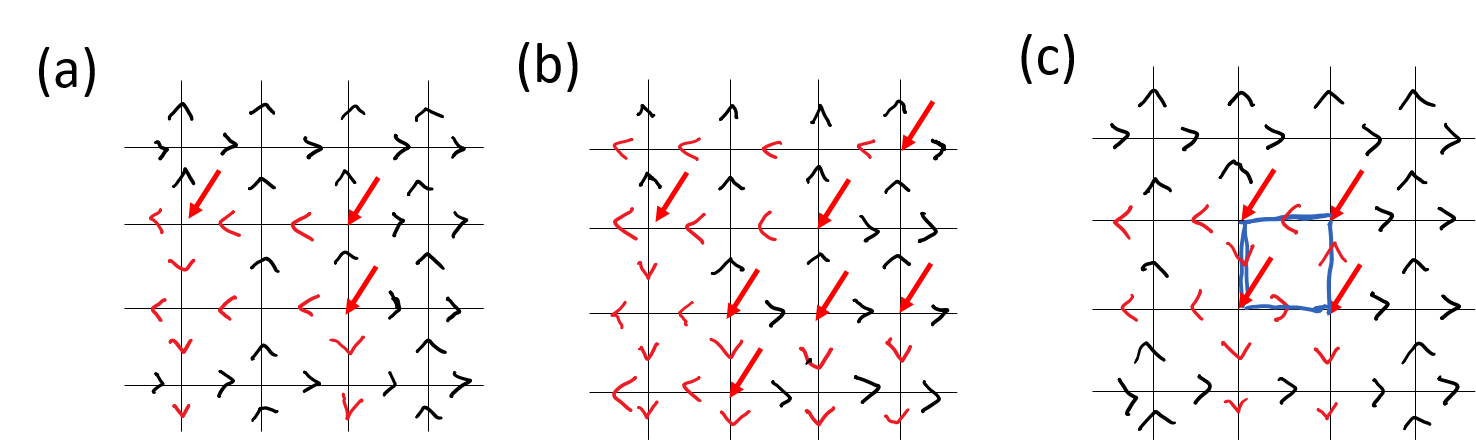}
\caption{(a) and (b) Arrow assignments that defines isometries from allowed configurations to the boundary. (c) An example of disallowed configurations for which no arrow assignment can be drawn without loop.
} \label{fig:code2}
\end{center}
\end{figure}

When there are $P$ sites on the boundary, any bulk region with more than $P/2$ sites cannot be mapped to the boundary isometrically, since the dimension of the Hilbert space cannot decrease under an isometry. For a bulk set with $n\leq P/2$ sites, an isometry may or may not exist depending on the configuration. If a collection of $n$ bulk sites $C=\left\{x_1,x_2,...,x_n\right\}$ can be mapped to the boundary isometrically, that guarantees that all operators acting on these $n$ sites are mapped to the boundary with their commutation relation preserved. In other words, these $n$ bulk sites can be viewed as having independent degrees of freedom, as long as all other sites remain in the ground state $|n=1\rangle$. For later convenience, we will call such a configuration an ``allowed configuration" and one that is not mapped isometrically a ``disallowed configuration". An example of allowed configuration is shown in Fig. \ref{fig:code2}.

\begin{figure} [h]
\begin{center}
\includegraphics[width=0.9\textwidth]{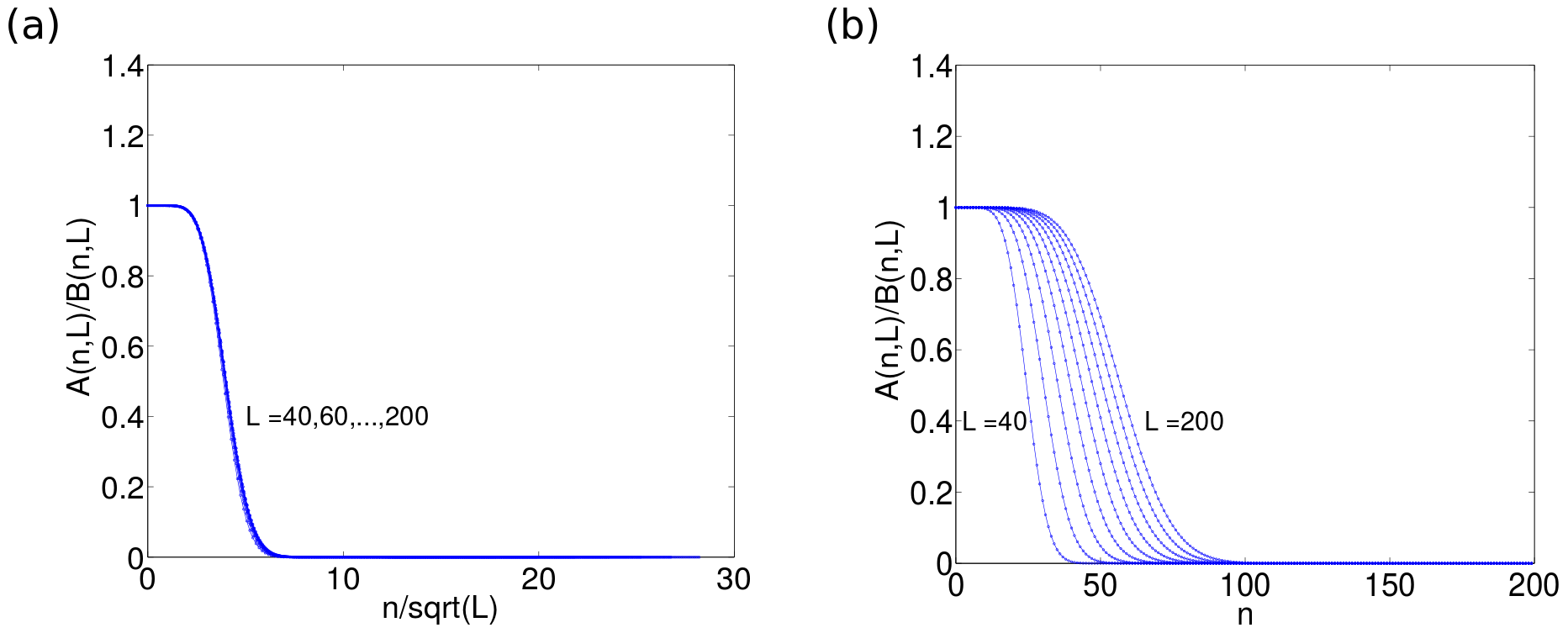}
\caption{(a)The percentage of allowed $n$-site configurations $A_n/B_n$ as a function of $n/\sqrt{L}$ on a $L\times L$ square lattice. The curves for different $L$ collapse to the same curve which demonstrates the $n/\sqrt{L}$ scaling behavior. As a comparison, (b) shows $A_n/B_n$ as a function of $n$ on a $L\times L$ square lattice.
} \label{fig:AoverB}
\end{center}
\end{figure}

Starting from the vacuum, any measurements that only access an allowed configuration will not be able to distinguish the bulk theory from an ordinary quantum many-body system with a factorizable $D^2$-dimensional Hilbert space at each site. A ``bulk observer" who measures multipoint functions in the bulk vacuum state would only discovery that the theory is not a local QFT when he/she measures an $n$-point functions on a disallowed $n$-site configuration. Therefore an important kinematic property of the bulk theory defined by the BHC is how many $n$-site configurations are allowed. For the bulk theory to be a good candidate of quantum gravity, one would like to see that for small enough $n$ almost every $n$-site configuration is allowed, so that the theory appears to be a local QFT for almost all low energy processes, and only deviates from a local QFT when one measures very complicated correlation functions (or equivalently, creating a high energy excitation above the vacuum).

For this purpose we would like to study the flat space square lattice as an example, and compute the number of $n$-site allowed configurations, denoted by $A^*_n$. When we consider allowed configurations that can be mapped isometrically by arrow drawing, we obtain a lower bound on $A^*_n$ since there may be other configurations that are allowed but cannot be shown by arrow drawing. In the following we denote this lower bound by $A_n$, which for the flat square lattice can be computed iteratively.

Since we compute $A_n$ iteratively with respect to the size of the
square lattice, we define $A(n,x,y)$ as the $A_n$ on a $x\times y$
rectangular lattice, and relate it to
$A(n^\prime,x^\prime,y^\prime)$ where $x^\prime, y^\prime,
n^\prime \leq x, y, n$. The key observation is that, if there is
an allowed configuration on an $x\times y$ rectangular lattice, it is
equivalent to say that, among the four boundaries of such
rectangular lattice, there exists at least one boundary along which
no bulk excitation or only one bulk excitation point exists. After excluding
that boundary from the $x\times y$ rectangular lattice,
the rest of the bulk points still form an allowed
configuration. From that observation, we can relate $A(n,x,y)$
to $A(n^\prime,x^\prime,y^\prime)$ with $n^\prime=
n,n-1,n-2,n-3,n-4$ and $x^\prime =x, x-1, x-2$, $y^\prime =y, y-1,
y-2$.  The details of evaluating $A(n,x,y)$ are described in Appendix
\ref{appendix2}. Then we take $x=y=L$, and calculate the ratio of
$A_n$ with the number of all $n$-site configurations
$B_n=\left(\begin{array}{c}L^2\\n\end{array}\right)$, which
measures the probability that an $n$-site configuration is
allowed. For an $L\times L$ square lattice, we find the following
scaling behavior of $A_n/B_n$:
\begin{eqnarray}
\frac{A_n}{B_n}=f\left(\frac{n}{\sqrt{L}}\right)
\end{eqnarray}
with $f( n / \sqrt{L} )$ a function that decreases gradually from $1$ at small $n$ to $0$ as $n$ approaches $2L$, as shown in Fig. \ref{fig:AoverB}. The scaling law tells us that for any small $\epsilon>0$, there exists a critical value $\alpha_\epsilon$, such that $\frac{A_n}{B_n}>1-\epsilon$ for all $n<\alpha_\epsilon \sqrt{L}$.  This result shows that disallowed configurations are very rare until the number of sites $n$ is comparable with $\sqrt{L}$. Since the $A_n$ obtained in our counting is a lower bound, the actual ratio $A^*_n/B_n$ for a specific BHC can be higher, which means that disallowed configurations may be even rarer.

Since $\sqrt{L}$ diverges at large $L$, any finite $n$ will be below the threshold for large enough $L$. In other words, in a big enough system with a square lattice, almost all $n$-point functions in the bulk look like those in a local QFT. When we create sufficiently many excitations, corresponding to changing the state on more than $\sqrt{L}$ sites, the deviation from QFT starts to be significant. This is the tensor network analog of the following process in an Einstein gravity theory: when a few particles are excited in the vacuum and interact with each other, the geometry can be considered to be a static background, and the process is well-described by a QFT. When more and more particles are sent to a bulk region, the gravitational back-reaction starts to be significant, which eventually leads to the formation of a black hole. In the BHC one can consider $n\sim \sqrt{L}$ as the ``energy scale" at which backreaction starts to be significant, and $n=2L$ as that of black hole formation. For different choices of tensors, the threshold $n_c$ above which the backreaction is significant may change, but for any BHC formed by \ideal tensors, $n_c$ always grows with $L$ at a rate equal to or greater than $\sqrt{L}$.

\section{Conclusion and discussion}

We have introduced the notion of a \ideal tensor and used networks of them to construct a holographic mapping we call the bidirectional holographic code (BHC). The BHC has various nontrivial properties which indicate its relevance to the understanding of holographic duality. The BHC defines a unitary mapping between the entire Hilbert space of the boundary theory and that of the bulk theory. The physical Hilbert space of the bulk theory is a subspace of the naive bulk Hilbert space, defined by a gauge invariance condition. Unlike an ordinary gauge symmetry, this gauge invariance condition implies that there are no local operators defined on the physical Hilbert space. However, the properties of the BHC guarantee that some direct product states in the bulk correspond to states on the boundary satisfying the RT formula for single boundary regions. Physically such states correspond to classical geometries in a quantum gravity theory. Nontrivial operators can be defined on some subset of the bulk points if the remainder of the bulk is known to be in such a classical state. These operators are the analogs of local operators acting on a given background geometry in a gravity theory. A single-site operator in the bulk defined in this sense can be mapped to multiple different boundary regions, in agreement with the ``error correction conditions" discussed in Ref. \cite{almheiri2014}. All bulk sites that can be mapped isometrically to a given boundary region form the causal wedge of the region. We studied the mapping of bulk multi-point operators to the boundary for a flat space $L \times L$ square lattice, and showed that below a critical number of sites $n_c\propto \sqrt{L}$, almost all $n$-site configurations in the bulk are mapped to the boundary isometrically. Consequently  the highly nonlocal bulk theory appears to have local degrees of freedom on each site, and it is difficult to observe the nonlocality until highly complicated correlation functions are measured. This is the key property which suggests that the bulk theory defined by the BHC is ``holographic" and has emergent local degrees of freedom at sub-AdS scale or even in flat space.

The construction of the BHC suggests many new questions. The BHC defines a mapping between the boundary theory and the bulk theory which does not directly describe the dynamics of either system. A boundary Hamiltonian $H$ can be mapped to the bulk physical Hilbert space to define the bulk Hamiltonian $H_b=MHM^\dagger$, which is generically non-local in the bulk degrees of freedom. Bulk locality in the low energy dynamics will only be preserved for very special boundary Hamiltonians. It is an open question how such Hamiltonians can be identified.

Another interesting problem is to determine the physical interpretation of a geometry with positive curvature. A BHC on a geometry with positive curvature is still well-defined, but the boundary states corresponding to classical geometries generically do not satisfy the RT formula. Do such geometries still have an interpretation in terms of entanglement? In the same spirit, for geometries such as de Sitter space that have no spatial boundary, a BHC naively defines just a single state in the bulk. What is the interpretation of geometry in this case? Although we do not have answers to these questions, the BHC approach at least provides a concrete starting point to begin investigating them. Other open questions include how to take the continuum limit of the BHC, and how to incorporate boundary states with global symmetries. We will leave these fascinating questions to future work.



\section*{Acknowledgments}

The authors thank Bartek Czech, Sepehr Nezami, Fernando Pastawski, Jamie Sully, Michael Walter and Beni Yoshida for enlightening conversations. ZY is supported by the David and Lucile Packard Foundation. PH was supported by the CIFAR, FQXi and the Simons Foundation. XLQ is supported by the National Science Foundation through the grant No. DMR-1151786, and also partially supported by the Templeton Foundation.

\appendix

\section{Pluperfect tensor search for stabilizer codes}\label{app:stabilizer}

There are multiple ways to describe quantum stabilizer codes. In
this section, we summarize the description we use in this work,
which is to map stabilizers to vectors over the integers mod $D$, and show
how to translate the three pluperfection conditions into this
language. For a more complete introduction to the the theory of stabilizer codes, the reader can consult Refs.~\cite{gottesman97,sarvepalli2007nonbinary}.

A qudit stabilizer code is a set of commuting operators
constructed out of $D$-dimensional on-site Pauli matrices $\{X_D,Z_D\}$, where $(X_D)_{mn} = \delta_{m,(n-1 \text{mod} D)}$, $(Z_D)_{m,n} = \delta_{m,n} e^{\frac{2m\pi i}{D}}$. It is straightforward to check that
\begin{equation}
   X_D Z_D= e^{ \frac{2\pi i}{D}} Z_D X_D.
\end{equation}
Every $k$-qudit stabilizer operator is mapped to a 2$k$-dimensional vector,
whose every element is an integer from $0$ to $D-1$. In the first
$k$ entries, the value of the $m$th element , $m\in[1,k]$,
represents how many times $X_D$ acts on the $m$th qudit; and from
the $k+1$ to the $2k$ entries, the value of the $m$th element ,
$m\in[k+1,2k]$, represents how many times $Z_D$ acts on qudit $m-k$.
For example, the $[403]_3$ code $\left\{S_1,S_2,S_3,S_4\right\}=\left\{ZZZI,~ZZ^{-1}IZ,~XXXI,~XX^{-1}IX\right\}$ (Eq.\ref{stabilizers}) can be represented as
\begin{equation}
  \label{eq:3}
  S\equiv\left(
  \begin{array}{cccccccc}
0& 0& 0& 0& 1& 1& 1& 0\\
0& 0& 0& 0& 1& 2& 0& 1\\
1& 1& 1& 0& 0& 0& 0& 0\\
1& 2& 0& 1& 0& 0& 0& 0
\end{array}\right)
\end{equation}
where the four row vectors $\{\vec{S}_1, \vec{S}_2, \vec{S}_3,
\vec{S}_4\}$ represent the four stabilizer operators. For simplicity, we choose $D$ prime so that $\mathbb{Z}_D$ is a finite field. The commutation relation is naturally expressed as a linear algebraic relation:
\begin{eqnarray}
  \label{eq:1}
  [S_i,S_j] = 0 &\Leftrightarrow& \vec{S}_i\cdot M \cdot \vec{S}_j^T=0 \\
\nonumber \text{where } M &=&\left(
\begin{array}{cc}
  0& I_{k\times k} \\
  - I_{k\times k} & 0
\end{array}
\right).
\end{eqnarray}

Finally, let's translate the three pluperfection conditions
into this new language.  As discussed in Section~\ref{sec:bhc}, if we start from a 4-leg perfect tensor stabilized by $\{S_1,S_2,S_3,S_4\}$, and apply operators that are direct product of unitary operators acting on each leg, then Condition \ref{cond2} is automatically satisfied. We define two operators $A$ and $B$, which are products of Pauli matrices corresponding to vectors $\vec{A}$ and $\vec{B}$ over $\mathbb{Z}_D$ respectively. The states $|nm\rangle=A^nB^m|0\rangle$ therefore satisfy Condition \ref{cond2} for any $n,m$ (with $|0\rangle$ the state stabilized by  $\{S_1,S_2,S_3,S_4\}$). To write down Condition \ref{cond3} we consider a generic single site operator $O_{a,b}= X^a Z^b$. Condition \ref{cond3} is equivalent to the statement that for all $a,b=0,1,2$, 
\begin{eqnarray}
\langle nm|O_{ab}|kl\rangle=\frac1D{\rm tr}(O)\delta_{nk}\delta_{ml}=\delta_{a0}\delta_{b0}\delta_{nk}\delta_{ml}.
\end{eqnarray}
Since
\begin{eqnarray}
\langle nm|O_{ab}|kl\rangle=e^{i\theta}\langle 0|A^{k-n}B^{l-m}O_{ab}|0\rangle,
\end{eqnarray}
with $\theta$ a phase obtained from permutation of $A^kB^l$ with $O_{ab}$, we see that condition \ref{cond3} requires the operators $A^{k-n}B^{l-m}O_{ab}$ to have zero expectation value in the state $|0\rangle$ unless $k=n,l=m$ and $a=0,~b=0$. This requires that the operators $A^{k-n}B^{l-m}O_{ab}$ each fail to commute with at least one of the stabilizers, unless $k=n,l=m$ and $a=0,b=0$ (in which case $A^{k-n}B^{l-m}O_{ab}=\mathbb{I}$ is the identity operator). This requirement can be expressed linear algebraically as:
\begin{eqnarray}
  \label{eq:2}
  S \cdot M \cdot \left(n\vec{A}+m\vec{B}+ a \vec{e}_i + b \vec{e}_{i+4}  \right)^T &=& 0, \\
  \text{only~if~}n=m=a=b&=&0\nonumber
\end{eqnarray}
where $\vec{e}_i,~i=1,2,\cdots,8 $ is the row vector with all entries $0$ except for the $i$th, which is 1, which represents the single site operator $X$ or $Z$ acting on the $i$th leg. The statement is required to hold for each $i$. Equivalently, the condition can be expressed as the requirement that for each $i$ the following $4\times 4$ matrix has full rank over $\mathbb{Z}_D$:
\begin{eqnarray}
L_i=S \cdot M \cdot \left[\vec{A}^T,\vec{B}^T, \vec{e}_i^T, \vec{e}_{i+4}^T  \right].
\end{eqnarray}
Here $\left[\vec{A}^T,\vec{B}^T, \vec{e}_i^T, \vec{e}_{i+4}^T
\right]$ denotes the $8\times 4$ matrix with the four vectors
$\vec{A}^T,\vec{B}^T, \vec{e}_i^T, \vec{e}_{i+4}^T $ as its
columns.

By generating some random vectors $\vec{A}$ and $\vec{B}$ and verifying these conditions, it is easy to find operators $A,B$ that define \ideal tensors.

\section{Dynamic programming algorithm bounding the number of allowed configurations}\label{appendix2}

In this appendix, we show explicitly how to calculate our lower bound $A(n,x,y)$ on the number of allowed
configurations on an $x\times y$ rectangular lattice. A sufficient
and necessary condition for being able to construct an isometry from the bulk
excited points to the boundary using only the properties of \ideal tensors is the existence of an arrow assignment to all contracted edges which satisfies the following conditions: a) for every vertex,
the number of outgoing edges is no less than that of the incoming edges;
b) there are no loops in the directed graph. These conditions
will be described below in an equivalent form which defines the
dynamic programming algorithm for calculating $A(n,x,y)$.

The conditions imply the existence of an isometry from the bulk
excited states to the boundary if and only if, starting from some
curves in the dual lattice, we can push the curves iteratively to
the boundary by a series of moves which satisfy: a) every move must
 include a vertex in the graph; b) if the vertex included is a
bulk excitation point, the length of the curve must increase; if it
is not, the length must not decrease; c) the region swept by the
curves contains all the bulk excitation points. These conditions
hold for general graphs.

\begin{figure}[h]
  \centering
\includegraphics[width=0.9\textwidth]{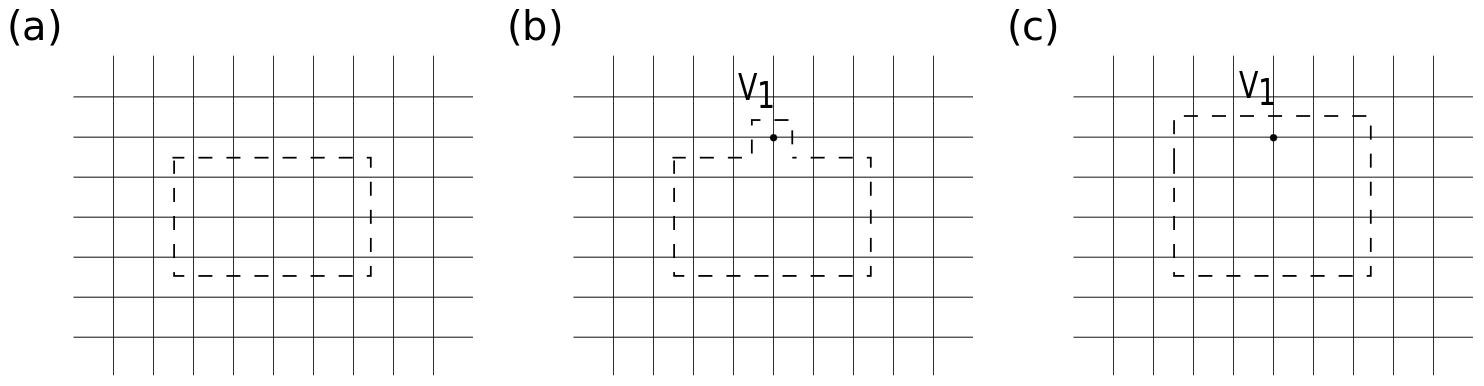}
\caption{The three steps used to grow an allowed rectangle while guaranteeing that the length of the curve be non-decreasing, as described in the main text.
}
 \label{UpdateRule}
\end{figure}

On a rectangular lattice, the calculation of $A(n,x,y)$ can be simplified by using recurrence relations. Since an allowed configuration on a rectangle must still be an allowed configuration when restricted to a smaller rectangular subregion, we only need to study the allowed moves that expands a smaller rectangle to a bigger one. It is easy to see that one edge of a rectangle can only be expanded by one step through the following process (see Fig. \ref{UpdateRule}): a) Start from a curve that is a rectangle; b) Increase the length of the curve by including another vertex $V_1$ right next to one of the edges of the rectangle ; c) include all the vertices one by one that are in the same row or column of with $V_1$ and that already have two edges contained in the curve, until the curve again forms a rectangle. If there is more than one bulk excitation point next to one edge, there is no allowed move that can include them and expand the rectangle.    

From the above update rules on the rectangular lattice, we
conclude that if there is an allowed configuration on an $x\times
y$ rectangular lattice, it is equivalent to say that, among the
four boundaries of the lattice, there exists at least one boundary
along which: (i) no bulk excitation or only one bulk excitation
point exists; (ii) excluding that boundary from the $x\times y$
rectangular lattice, the rest of the bulk points still form an
allowed configuration. Therefore, we can continue this process
until the rectangle boundaries shrink to a point. This process
relates the number of allowed configurations on the $x\times y$
square lattice $A(n,x,y)$ to the number on a smaller rectangle
lattice.  In fact, there are at most four ways to relate an
allowed configurations on an $x\times y$ rectangular lattice to a
smaller rectangular lattice, in particular, to eliminate the up,
down, left, or right boundary. Thus we obtain
\begin{equation}
  A(n,x,y) = F_1(n,x,y) - F_2(n,x,y) + F_3(n,x,y) - F_4(n,x,y)\label{iterationeq1}
\end{equation}
where $F_1(n,x,y)$ counts the number of allowed configurations on
the $x\times y$ rectangular lattice that can be reduced to an allowed
configuration on a smaller lattice by eliminating one of the up,
down, left, or right boundary, while the eliminated boundary contains
at most one bulk excited point. However, some allowed
configurations on $x\times y$ can reduce to an allowed
configurations on a smaller lattice via at least two
approaches. These are counted at least twice in $F_1(n,x,y)$. To
eliminate such over counting, we include $F_2(n,x,y)$,
$F_3(n,x,y)$, $F_4(n,x,y)$, which calculate the number of allowed
configurations on the $x\times y$ rectangular lattice that can relate to
allowed configurations on a smaller lattice via at least 2, 3, 4
approaches. $F_{1,2,3,4}(n,x,y)$ can be determined by $A(m,x',y')$ on smaller rectangles with fewer points. For example, if there are $A(n-1,x-1,y)$ number of allowed configurations on a rectangle with size $(x-1)\times y$, by adding one bulk input site at the left edge of it we get $yA(n-1,x-1,y)$ allowed configurations in the rectangle with size $x\times y$. The same number of configurations is obtained by adding a site to the right edge. By considering all such expansion processes, we obtain the following recurrence relations:
\begin{eqnarray}
  \nonumber F_1(n,x,y) &=& 2 \left[
    y\cdot A(n - 1, x - 1, y) + x\cdot A(n -1, x, y - 1) + A(n, x - 1, y) + A(n, x, y - 1)
  \right] \\
  \nonumber F_2(n,x,y) &=& 4 \Big[
  (x - 1) \cdot (y - 1)\cdot A(n - 2, x - 1, y - 1) + (x + y - 1)\cdot A(n - 1, x - 1, y - 1) \\
  \nonumber &&		+ A(n, x - 1, y - 1)
  \Big] + \Big[
  (x^2 \cdot A(n - 2, x, y - 2) + 2x \cdot A(n - 1, x, y - 2) \\
  \nonumber &&	+ A(n, x, y - 2))	+(x\leftrightarrow y)  \Big] \\
  \nonumber F_3(n,x,y) &=& \Big\{
  2 \Big[
  (x - 1)^2 \cdot (y - 2) \cdot A(n-3, x - 1, y - 2)
  + (x - 1) \cdot (x + 2 y - 3) \cdot A(n - 2, x - 1, y - 2) \\
  \nonumber &&+ (2 x +  y - 2) \cdot A(n - 1, x - 1, y - 2) + A(n, x - 1, y - 2)
  \Big]
  + (x\leftrightarrow y)
  \Big\} \\
  \nonumber F_4(n,x,y) &=&
  (x - 2)^2 \cdot (y - 2)^2 \cdot A(n - 4, x - 2, y - 2) \\
  \nonumber &&  + (x - 2) \cdot (y - 2)\cdot  (2 x + 2 y - 4)\cdot A(n - 3, x - 2, y - 2) \\
   \nonumber &&+ \left[
    x^2 + y^2 + 4 x y - 8 (x + y) + 10
  \right] A(n - 2,  x - 2, y - 2) \\
    &&+ 2 (x + y - 2)A(n - 1, x - 2, y - 2) + A(n, x - 2, y - 2)\label{iterationeq2}
\end{eqnarray}
Therefore Eq. (\ref{iterationeq1}) and (\ref{iterationeq2}) define the recurrence relations satisfied by $A(n,x,y)$. The initial conditions are
\begin{eqnarray}
  \label{eq:4}
 \nonumber && A(n,x,y) = 0,~\textnormal{if}~ n<0 ~ \textnormal{or} ~ x<0 ~ \textnormal{or}~ y<0 \\
\nonumber && A(0,0,0) = A(0,0,1) =A(0,1,0) = A(0,1,1) =A(1,1,1) = 1\\
 && A(1,0,0) = A(1,0,1) = A(1,1,0) =0. \label{initalcond}
\end{eqnarray}
These initial conditions are fixed by
\begin{eqnarray*}
  A(0, i, j) = 1,~ \forall ~ i>0, ~ j>0,~  A(1,1,i) = i,~ \forall ~ i>0
\end{eqnarray*}
and it can be easily checked that
\begin{equation*}
  A(n,i,j) = \binom{i\times j}{n},~n=0,1,2,3
\end{equation*}
which means all configurations with up to $3$ excitation points
are allowed.  We can then determine $A(n,x,y)$ for any $n,x,y$
from Eq. (\ref{iterationeq1}), (\ref{iterationeq2}) and
(\ref{initalcond}).

\bibliographystyle{unsrt}
\bibliography{ehm}

\end{document}